\documentclass[12pt,a4paper]{article}

\usepackage{arxiv}
\usepackage[utf8]{inputenc}
\usepackage[T1]{fontenc}

\usepackage{graphicx}
\usepackage{xcolor}

\usepackage{booktabs}
\usepackage{longtable}
\usepackage{array}
\usepackage{calc}
\usepackage{pdflscape}
\usepackage{xurl}

\usepackage{amsmath,amssymb}
\usepackage[numbers]{natbib}
\usepackage{caption}
\usepackage[hidelinks]{hyperref}
\hypersetup{
  pdftitle={From Policing Provenance to Designing for Evidence: Testing Four Propositions on Generative AI Against Research and Institutional Policy},
  pdfauthor={Samar Ansari},
  pdfkeywords={generative AI; assessment validity; higher education assessment; academic integrity; assessment design; AI detection; comparative policy analysis}
}

\title{\textbf{From Policing Provenance to Designing for Evidence:\\
Testing Four Propositions on Generative AI Against Research and Institutional Policy}}

\author{
	Samar Ansari\\
	School of Computing and Engineering Sciences\\
	University of Chester\\
	Chester, CH1 4BJ, United Kingdom \\
	\texttt{m.ansari@chester.ac.uk}
}

\date{\today}

\begin{document}

\maketitle

\begin{abstract}
\noindent Generative AI has made it hard to establish whether a student produced a piece of assessed work. Building on the argument that validity matters more than cheating, this paper tests four propositions drawn from a strongly stated position: that redesign is more sustainable than detecting or banning AI; that asking students whether and how much AI they used yields low-value information; that unaided ability should be tested directly in controlled conditions; and that the same logic applies to peer review. The propositions are tested against two purposively sampled, coded corpora: 41 research and sector sources and 55 university and sector policy documents. Coding was by a single coder and the policy sample is UK-weighted, so counts describe the sample, not the sector. The propositions hold unevenly. Redesign over detection is well supported, but only as redesign plus assurance: 4 of 55 documents prohibit AI by default, though about 10 of the 55 are publisher or editorial policies that govern authors rather than student assessment; of the 23 that take a position on detection, 16 reject detectors and one requires them. Testing unaided ability is supported as one lane within a programme, not as a return to invigilation. The disclosure proposition is complicated: institutions treat disclosure as valuable, yet the limited evidence suggests many students do not complete required declarations. The peer-review extension is plausible but under-examined. The field is converging on graded permission scales and two-lane designs, and 31 of 55 documents define named categories of use. From these findings the paper builds a nine-step decision chain that extends outcome-first design logic: the role of AI is decided after the evidence an outcome requires. It also separates design moves that add temporary friction from those that provide assurance. A worked example and the full policy matrix are included.
\end{abstract}


\section{Introduction}

\subsection{The problem is validity, not provenance}

Within three years of the public release of large language models, the use of generative AI in higher education assessment has moved from a marginal concern to a near-universal condition. Recent United Kingdom student surveys report that almost all students now use these tools and that the great majority use them for assessed work \citep{StudentGenerativeAISurvey2025,StudentGenerativeAISurvey2026}. The first institutional reflex, to treat such use as misconduct and to detect it, has not held up: controlled testing shows that AI-text detectors are neither accurate nor reliable \citep{weber2023testing} and that they misclassify the writing of non-native English speakers at high rates \citep{liang2023gpt}. The practical result is that the question most assessment procedures are built to answer, namely whether a human produced a given piece of work, can no longer be answered with confidence.

Following \citet{dawson2024validity}, who argue that validity matters more than cheating, this paper treats the more important problem as validity rather than provenance. When a universally available tool can produce the graded artefact, a mark no longer certifies that the student performed the intended intellectual work. The threat is sharpened by evidence that AI can raise measured performance while lowering actual learning: a randomised trial of nearly 1,000 secondary-school mathematics students \citep{bastani2025guardrails} found large gains in practised performance alongside lower scores on an unaided exam in the same session for students given unguided AI access. The challenge for higher education, then, is not to catch AI use but to design assessment so that it still establishes what it claims to establish about a student. This reframing, from policing provenance to designing for evidence, was argued conceptually by \citet{dawson2024validity}, who treat inappropriate AI use as a property of the assessment rather than of the student. This paper takes that position as its organising claim, tests it against policy and literature evidence, and turns it into a design procedure.

\subsection{The anchoring position and four propositions}

The study is organised around a strongly stated opinion piece, referred to here as the anchoring post \citep{terbeck2026assume}, which puts this reframing in its strongest form. The post argues that unless an assessment explicitly forbids external help, we should assume generative AI has been used, possibly for the whole task; that asking whether and how much AI was used produces little of value and encourages a culture of confession; that where we genuinely need to know what a student can do unaided we should test that directly through controlled conditions such as invigilated examinations, vivas and observed practical work; and that the same logic extends to scholarly peer review.

The post is an opinion piece, not peer-reviewed research, so this paper treats it as an analytical foil rather than a position to defend. Its claims are restated as four propositions that published evidence and policy can be used to support or challenge:

\begin{itemize}
\setlength{\itemsep}{2pt}
\item \textbf{P1.} Where AI can perform the intended intellectual work of a task, redesigning the assessment is more sustainable than detecting or banning AI.
\item \textbf{P2.} Disclosure questions, asking whether and how much AI was used, produce low-value information.
\item \textbf{P3.} Unaided human ability should be tested directly in controlled conditions.
\item \textbf{P4.} The same logic applies to scholarly peer review: a review should be judged by its quality, not by how much a human typed.
\end{itemize}

Framing the post this way turns a provocative argument into testable claims, and it allows the paper to report where the evidence agrees with the post, where it is mixed, and where it pushes back, rather than simply endorsing or rejecting a slogan.

\subsection{The gap and the contribution}

Reviews of generative AI in higher education are now numerous, and many repeat student-perception surveys. Recent work has begun to offer design logics for assessment under generative AI, including outcome-derived rules for AI use \citep{klapproth2026redesigning}, assurance typologies \citep{elkhodr2026assurance}, discipline-specific design matrices \citep{ahangama2026designing} and multi-pillar frameworks \citep{nikolic2026designing}. Some studies analyse scholarly and institutional documents together \citep{ahangama2026designing}, but what remains rare is a test of a stated position against the research literature and institutional policy separately. This paper makes three contributions. First, it tests four propositions drawn from a strongly stated position against two coded corpora, one of research and sector sources and one of policy documents. Second, it compares how universities and sector bodies are responding, coded against two typologies: one for the institutional response paradigm and one for the assessment logic a design assumes. Third, it extends existing outcome-first design logic into a nine-step decision chain that adds explicit steps for AI capability, AI as a graduate capability, the assurance mechanism and its costs, and demonstrates the chain through a worked example. The capability of AI itself is treated through a cognitive lens that is used, and then explicitly bounded, so that the design logic rests on assuring a student's reasoning rather than on hunting for tasks that AI cannot yet perform.

\subsection{Aim and research questions}

The aim of the study is to test four propositions about generative AI and assessment against research and institutional policy evidence. To do so, it maps how generative AI challenges the validity of higher education assessment, how institutions and sector bodies are responding, and what principles for assessment design emerge from existing evidence. Five research questions structure the evidence, and the Discussion then returns to the propositions:

\begin{itemize}
\setlength{\itemsep}{2pt}
\item \textbf{RQ1.} What problems does student generative AI use create for the validity, integrity and equity of higher education assessment?
\item \textbf{RQ2.} How are universities and sector bodies responding, and how do responses vary by country and institution type?
\item \textbf{RQ3.} What assessment designs and frameworks keep assessment valid when AI use is presumed?
\item \textbf{RQ4.} What trade-offs do these approaches involve for learning, integrity, equity, workload, privacy and validity?
\item \textbf{RQ5.} What plausible future directions emerge, and under what conditions?
\end{itemize}

\subsection{Approach and structure}

The study is a secondary-research synthesis of two corpora: a body of research and sector literature, and a sample of institutional and sector policy documents. Sources were gathered pragmatically, through targeted searching, citation chasing and purposive selection of institutions for variation, rather than through an exhaustive systematic search with a logged screening tally. This is therefore a pragmatic, exploratory review rather than a systematic review, and the sample is purposive and weighted toward the United Kingdom and the Anglophone world; the methodology section states these limits in full, and they bound every count the paper reports.

The remainder of the paper proceeds as follows. The methodology describes the two corpora, the coding typologies and the appraisal approach. A conceptual framework then sets out the design logic, the capability lens and the worked example. Five findings sections address RQ1 to RQ5 in turn: the problems for validity, integrity and equity; the institutional and sector responses; the assessment-design models; the trade-offs; and the future directions, framed as scenarios. A discussion returns to propositions P1 to P4, and a conclusion states the contribution and its limits.

\section{Methodology}

\subsection{Design}

This study is a secondary-research synthesis combining a review of research and sector literature with a comparative analysis of institutional and sector policy. It is best described as a pragmatic, exploratory review rather than a systematic review: it does not claim exhaustive coverage, and it was not conducted against a pre-registered protocol or a logged screening tally. The aim was breadth of variation across responses and designs, not a complete census, and the reporting below is written to make that scope, and its limits, explicit.

\subsection{The two corpora}

The evidence sits in two corpora. The literature corpus holds 41 sources, spanning peer-reviewed studies, reviews, conceptual papers and sector reports. Six of these 41 are methodology references \citep{arksey2005scoping,levac2010scoping,tricco2018prisma,braun2006using,thomas2008methods,bowen2009document}, on scoping and thematic-synthesis technique, that the study draws on for its own method rather than reviews as evidence; they are retained in the corpus count but carry no proposition coding, and they account for six of the nine literature sources coded not applicable on the propositions. The policy corpus holds 55 documents: university policies and guidance, sector and quality-assurance body guidance, and professional-body statements. Three of the 55 are research publications rather than single-institution policy statements: two are systematic mappings of UK policy (Tao et al.; Atkinson-Toal and Guo) and one is a framework paper (the AI Assessment Scale); Section~\ref{sec:rq2} explains how the two mappings are read. Table~\ref{tab:corpus} gives the composition of both. Three features matter for interpretation. First, the policy corpus mixes document types, so counts over it pool university policies with sector and publisher documents. Second, the policy sample is weighted toward the United Kingdom and the Anglophone world, so cross-national comparisons are illustrative rather than representative. Third, counts are over documents, not institutions: a few bodies contribute several documents each (for example nine from King's College London), so the frequencies describe the document set rather than a population of universities.

\begin{table}[htbp]\centering\footnotesize
\caption{Composition of the two corpora.}\label{tab:corpus}
\begin{tabular}{@{}p{7.2cm} r r@{}}\toprule
\textbf{Category} & \textbf{n} & \textbf{\%} \\\midrule
\multicolumn{3}{@{}l}{\textit{Literature corpus by source type (N=41)}}\\
\quad Conceptual/Theoretical & 20 & 49\% \\
\quad Empirical study & 11 & 27\% \\
\quad Systematic/Scoping review & 5 & 12\% \\
\quad Commentary/Editorial & 3 & 7\% \\
\quad Report & 2 & 5\% \\
\midrule
\multicolumn{3}{@{}l}{\textit{Policy corpus by source type (N=55)}}\\
\quad University guidance & 28 & 51\% \\
\quad Other grey literature & 12 & 22\% \\
\quad University policy & 8 & 15\% \\
\quad Sector/QA body guidance & 4 & 7\% \\
\quad Professional body & 3 & 5\% \\
\midrule
\multicolumn{3}{@{}l}{\textit{Policy corpus by jurisdiction (N=55)}}\\
\quad UK & 26 & 47\% \\
\quad International / Global & 10 & 18\% \\
\quad USA & 4 & 7\% \\
\quad Australia & 3 & 5\% \\
\quad Canada & 2 & 4\% \\
\quad Hong Kong & 2 & 4\% \\
\quad South Africa & 2 & 4\% \\
\quad Belgium & 1 & 2\% \\
\quad Sweden & 1 & 2\% \\
\quad Ireland & 1 & 2\% \\
\quad Singapore & 1 & 2\% \\
\quad UAE & 1 & 2\% \\
\quad Switzerland & 1 & 2\% \\
\bottomrule\end{tabular}\end{table}

\subsection{Source identification}

Sources were gathered pragmatically. Literature was located through targeted keyword searching, citation chasing from key reviews, and three compiled policy datasets that were used as leads to individual institutions \citep{JIAO2024LANDSCAPE,MOORHOUSE2023ASSESS,AGGA2025DATASET}. Institutional policies were identified by purposive selection of universities chosen to vary by region, institution type and stance, and were read from each institution's own published pages. Because collection was pragmatic, the corpora may carry selection effects, including an over-representation of authors and institutions that actively publish on assessment redesign; this is treated as a limitation rather than corrected. Where policy pages gave a date or version, it was recorded; readers are reminded that policy pages change, and archiving each page at the time of access is recommended for replication. A small number of directly relevant works, some published while the corpora were being finalised \citep[for example][]{dawson2024validity,klapproth2026redesigning,elkhodr2026assurance,ahangama2026designing}, are discussed in the framing and discussion but were not added to the coded corpora, so the counts reported here are unaffected by them.

\subsection{Scope boundaries}

Inclusion was limited to higher education, to generative AI rather than older text-matching technology, and to material from 2022 onward, with foundational assessment theory admitted as background. Sources were in English, with one exception: the FDUSP report \citep{USPFDUSP2024}, a Portuguese-language document from which specific figures were read directly. This near-exclusive use of English is otherwise a substantive limitation given the equity and access themes the paper reports. Sources that concerned only school-level education, or that were purely technical with no assessment relevance, were excluded.

\subsection{Data extraction and coding}

Each source was extracted into a fixed schema capturing bibliographic fields, study characteristics, key findings, and assessment, detection and equity implications. Extraction was assisted by a large language model operating over the full text of each source under a structured prompt, with the output reviewed and corrected by the author; this AI-assisted step is disclosed here because it is itself a methodological choice with error modes, and a data-cleaning pass corrected several mis-classified and mislabelled records before analysis. Generative AI tools were also used earlier, at the planning and drafting stage, to help generate initial versions of the research questions, the first proposition (P1), the Typology~2 table, the structure of the decision chain in Section~\ref{sec:framework}, the worked example in Section~3.6, and several of the future scenarios in Section~8. The author then developed each of these, grounded them in the cited literature, revised them, and checked them against the sources. The coding was itself AI-assisted in the same way: the extraction model proposed an initial code for each field, and the author reviewed and finalised every code against the source. The analysis and the conclusions are the author's own. This use is disclosed in full because transparency about AI assistance in study design is consistent with the argument the paper makes about assessment. Every source was additionally coded against two typologies and the propositions. Typology~1 classifies an institutional response as detect-and-punish, disclose-and-declare, redesign, or integrate. Typology~2 classifies an assessment by the role it assigns to AI, as secure, controlled, open, AI-integrated, or hybrid, as set out in the conceptual framework. In both typologies a document is coded \emph{Multiple} when it promotes more than one category, for example both secure and open assessment, and \emph{Unclear} when it states none; these two cells are descriptive residuals and are not read as endorsements of a single model. Each source was also coded for which of the four propositions it bears on and for the direction of that bearing, as supports, mixed, challenges or neutral.

\subsection{Quality appraisal}

Empirical sources were appraised with the Mixed Methods Appraisal Tool (MMAT) \citep{hong2018mmat} and grey literature and policy documents with the AACODS checklist (Authority, Accuracy, Coverage, Objectivity, Date, Significance) \citep{tyndall2010aacods}, each yielding an overall rating that is retained in the extraction and used to weight claims in the synthesis. Conceptual and theoretical sources, which are not amenable to either instrument, were marked as such and weighted by reasoned judgement rather than a formal score.

\subsection{Analysis}

Three analytic moves were combined. Thematic synthesis in the manner of \citet{thomas2008methods} was used for the issues, design and future questions (RQ1, RQ3, RQ5), coding findings into descriptive themes and then into the analytic themes reported in those sections. A comparable recent study used thematic analysis of 80 publicly available scholarly and institutional documents to derive an assessment framework for ICT education \citep{ahangama2026designing}. The present study differs in keeping research and policy sources in separate corpora, coding policy documents against fixed typologies, and testing stated propositions rather than deriving themes inductively. A comparative policy matrix was built for the responses question (RQ2), arraying each of the 55 documents against stance, paradigm, assessment logic, detection position, defined categories and proposition direction; the full matrix is given in Appendix Table~\ref{tab:matrix}. Frequency counts of the coded categories, reported in Tables~\ref{tab:polcounts} and~\ref{tab:litcounts}, summarise the distributions that the findings draw on; the full per-document policy codings are itemised in Appendix Table~\ref{tab:matrix} and the full per-source literature codings in Appendix Table~\ref{tab:litmatrix}. Throughout, counts are treated as descriptive of the sample, and the proposition mapping is read as the shape of the evidence rather than as a measurement. Many sources bear on more than one research question, so the per-RQ source counts reported in the findings sum to more than the corpus totals of 41 and 55.

\begin{table}[htbp]\centering\footnotesize
\caption{Coded distributions across the policy corpus.}\label{tab:polcounts}
\begin{tabular}{@{}p{7.6cm} r r@{}}\toprule
\textbf{Dimension and category} & \textbf{n} & \textbf{\%} \\\midrule
\multicolumn{3}{@{}l}{\textit{Response paradigm (Typology 1)}}\\
\quad Multiple & 20 & 36\% \\
\quad Disclose and declare & 17 & 31\% \\
\quad Redesign & 7 & 13\% \\
\quad Integrate & 7 & 13\% \\
\quad Unclear & 2 & 4\% \\
\quad Detect and punish & 2 & 4\% \\
\midrule
\multicolumn{3}{@{}l}{\textit{Assessment logic promoted (Typology 2)}}\\
\quad Multiple & 31 & 56\% \\
\quad Unclear & 11 & 20\% \\
\quad Hybrid & 7 & 13\% \\
\quad Controlled & 2 & 4\% \\
\quad Secure & 2 & 4\% \\
\quad AI-integrated & 1 & 2\% \\
\quad Open & 1 & 2\% \\
\midrule
\multicolumn{3}{@{}l}{\textit{Default stance}}\\
\quad Permit with limits & 24 & 44\% \\
\quad Devolved to staff & 14 & 25\% \\
\quad Unclear & 6 & 11\% \\
\quad Integrate/AI-expected & 4 & 7\% \\
\quad Prohibit & 4 & 7\% \\
\quad Disclosure-based & 3 & 5\% \\
\midrule
\multicolumn{3}{@{}l}{\textit{Detection position (bucketed)}}\\
\quad Silent & 32 & 58\% \\
\quad Avoid/ban & 16 & 29\% \\
\quad Indicator only & 5 & 9\% \\
\quad Lockdown/prevent & 1 & 2\% \\
\quad Require/use & 1 & 2\% \\
\midrule
\multicolumn{3}{@{}l}{\textit{Proposition direction}}\\
\quad Supports & 36 & 65\% \\
\quad Mixed & 14 & 25\% \\
\quad NA & 2 & 4\% \\
\quad Challenges & 2 & 4\% \\
\quad Neutral & 1 & 2\% \\
\bottomrule\end{tabular}\end{table}

\begin{table}[htbp]\centering\footnotesize
\caption{Coded distributions across the literature corpus.}\label{tab:litcounts}
\begin{tabular}{@{}p{7.6cm} r r@{}}\toprule
\textbf{Dimension and category} & \textbf{n} & \textbf{\%} \\\midrule
\multicolumn{3}{@{}l}{\textit{Response paradigm (Typology 1)}}\\
\quad Multiple & 18 & 44\% \\
\quad NA & 12 & 29\% \\
\quad Redesign & 6 & 15\% \\
\quad Integrate & 3 & 7\% \\
\quad Detect and punish & 2 & 5\% \\
\midrule
\multicolumn{3}{@{}l}{\textit{Assessment logic (Typology 2)}}\\
\quad Multiple & 14 & 34\% \\
\quad NA & 12 & 29\% \\
\quad Hybrid & 8 & 20\% \\
\quad Unclear & 3 & 7\% \\
\quad Open & 3 & 7\% \\
\quad Secure & 1 & 2\% \\
\midrule
\multicolumn{3}{@{}l}{\textit{AI capability zone}}\\
\quad NA & 18 & 44\% \\
\quad Mixed & 17 & 41\% \\
\quad AI-strong & 5 & 12\% \\
\quad AI-weaker & 1 & 2\% \\
\midrule
\multicolumn{3}{@{}l}{\textit{Proposition direction}}\\
\quad Supports & 22 & 54\% \\
\quad NA & 9 & 22\% \\
\quad Mixed & 8 & 20\% \\
\quad Neutral & 2 & 5\% \\
\bottomrule\end{tabular}\end{table}

\subsection{Tools}

Reference management, extraction spreadsheets and the coding typologies were maintained in standard tools, and the analysis tables were produced by scripted processing of the extraction files. Two counts reported in the findings, the number of documents carrying a volatility note (32 of 55) and the number invoking oral or process verification (21 of 55), are taken from the full policy extraction records rather than from the columns printed in the appendix matrix. Data availability: the comparative policy matrix (Appendix Table~\ref{tab:matrix}) and the literature coding table (Appendix Table~\ref{tab:litmatrix}) itemise both corpora in full, and the complete extraction records for both corpora, including each document's source link, date and version, are provided as supplementary files.

\subsection{Limitations of method}

Five limitations bound every result. First, the review is pragmatic and not systematic, so the corpora are not exhaustive and carry possible selection effects. Second, the typology, capability and proposition codings were proposed by the extraction model and reviewed and finalised by a single author; they await independent second-coding and inter-coder agreement, and should be read as indicative. The direction coding in particular assigns one overall direction to each document, not a separate code for each of the four propositions, so the aggregate direction counts are a document-level summary rather than per-proposition tallies. Third, extraction was AI-assisted, which introduces its own error modes despite human review, so individual cell values should be verified against sources before being quoted. Fourth, the policy corpus is UK-weighted and the review is almost entirely English-language (one Portuguese-language report excepted), so national and global claims are illustrative. Fifth, the policy documents record declared policy, not observed practice, so the study cannot speak to what staff and students actually do. These limitations are revisited in the discussion and shape the strength of every claim the paper makes.

\section{A conceptual framework: from provenance to evidence}
\label{sec:framework}

\subsection{The shift in question}

The framework developed here rests on a single move: it replaces the question that dominates current practice with a different one. The common question is forensic, "did a human produce this work," and it is the question that detection, disclosure statements and misconduct procedures all try to answer. The findings in RQ1 show that this question is now largely unanswerable, because detection is unreliable \citep{weber2023testing,liang2023gpt} and use is near-universal \citep{StudentGenerativeAISurvey2025,StudentGenerativeAISurvey2026}. The framework therefore asks instead a question of validity: what is this assessment meant to establish about the student, and does it still establish it when generative AI is available. Framing assessment this way turns the problem from policing provenance into designing for evidence, and it is the organising idea of the paper. \citet{dawson2024validity} make the same shift: the challenge generative AI poses is ``not one of student morality but one of assessment design''.

\subsection{Two classifications the framework builds on}

The framework uses the two typologies developed in the methodology and applied in the findings. Typology~1 classifies an institutional response as detect-and-punish, disclose-and-declare, redesign, or integrate, and locates where a policy sits. Typology~2, set out in Table~\ref{tab:logics}, classifies an assessment by the role it assigns to AI and by what it is therefore able to establish. The value of Typology~2 for design is that it makes the target explicit: a secure task establishes unaided ability, an open task establishes the quality of work produced with available tools, and these are different claims about a student that should not be confused.

\begin{table}[htbp]
\centering
\footnotesize
\caption{Typology~2: assessment logics by the role assigned to AI.}
\label{tab:logics}
\begin{tabular}{@{}p{2.1cm} p{2.4cm} p{4.4cm} p{4.6cm}@{}}
\toprule
\textbf{Logic} & \textbf{AI position} & \textbf{Typical form} & \textbf{What it establishes} \\
\midrule
Secure & AI excluded & Invigilated exam, viva, observed practical & What the student can do unaided \\
Controlled & Limited AI & Task with defined permitted uses & A human capability plus constrained assistance \\
Open & AI permitted & Essay, report, project & Quality of output produced with available tools \\
AI-integrated & AI required & AI critique, AI-assisted professional task & Ability to use, interrogate and evaluate AI \\
Hybrid & Both & Open task plus oral defence or process portfolio & Product quality plus evidence of individual understanding \\
\bottomrule
\end{tabular}
\end{table}

\subsection{The capability lens, and why it is a lens not a wall}

To decide whether AI can already perform an assessment's intended work, the framework borrows a cognitive taxonomy from the limits-of-LLMs literature, then immediately marks its limits. Following Peirce's three modes of inference, large language models are strong at induction, the recognition of patterns in data, and increasingly capable at deduction, the derivation of conclusions within a given frame; they are weakest at abduction, the generation of genuinely new explanatory ideas \citep{zahavy2026llms}. This paper states this in a deliberately softened form: \citet{zahavy2026llms} argues that current models are structurally incapable of the abductive step, whereas this paper treats abduction as the mode they are weakest at, because its argument is that the capability boundary moves rather than that it is fixed. Related work makes the narrower claim that these systems separate formal linguistic fluency from functional reasoning \citep{mahowald2024dissociating}, building on older arguments about symbol grounding \citep{harnad1990symbol} and about models that manipulate form without meaning \citep{bender2021stochastic}.

This lens is useful for one judgement: work that is pattern-based or derivable over public text sits in the zone AI handles well, while work grounded in a student's own data, context and live reasoning, or requiring novel framing, sits in a zone it handles less well. The lens must not be mistaken for a wall. Three limits apply. First, the boundary moves, so a task that resists AI today may not next year, which means designing around a current weakness is friction, not security. Second, the frontier example that motivates the taxonomy concerns scientific discovery, not coursework, and the RQ1 and RQ3 evidence shows AI already matching students on many higher-order and authentic tasks; the capability gap at the level of ordinary assessment is small and closing. Third, and consequently, the durable basis for design is not a search for tasks AI cannot do but the assurance of the individual student's reasoning. The lens therefore informs step~2 of the decision chain below, while the chain's weight rests on assurance.

\subsection{The assessment-design framework as a decision chain}

The framework is a sequence of questions for any assessment. It is a synthesis of the design principles found across the corpus, from authentic assessment \citep{wiggins1990case} and constructive alignment \citep{biggs1996enhancing} to assessment security \citep{dawson2020defending}, programmatic assessment \citep{schuwirth2011programmatic} and structured assessment-design decision frameworks \citep{bearman2016support}, organised so that the decision about AI follows from the decision about evidence rather than preceding it.

The principle that the rule for AI use should follow from the learning outcome is not new. \citet{klapproth2026redesigning} states it as a maxim and derives four options (AI prohibited, permitted, permitted with documentation, or required) from the intended outcome, and university guidance at the Australian National University \citep{anu2024rethinking} and Northeastern University \citep{northeastern2026assessment} sets out the same order. The chain below extends this ordering in four ways: it asks whether AI can already do the work and for how long (step~2); it asks whether using AI is itself part of the capability being certified (step~4); it requires the designer to name the assurance mechanism (step~8); and it requires the designer to count the costs of that mechanism (step~9). The chain keeps \emph{limited} at step~7 rather than Klapproth's \emph{permitted with documentation}, because \emph{limited} names the degree of permitted assistance, which Typology~2 already encodes as the controlled category, whereas documentation is one assurance mechanism among several and is decided at step~8.

\begin{enumerate}
\setlength{\itemsep}{2pt}
\item What is the intended learning outcome of this assessment?
\item Can generative AI already perform that intended work, and for how long is that answer likely to hold?
\item Must the student demonstrate this capability without AI?
\item Is using AI itself part of the graduate capability being certified?
\item What evidence of the student's own thinking is required?
\item What assessment format actually provides that evidence?
\item What role should AI then have: prohibited, limited, permitted, or required?
\item How is learning assured: through the product, the process, oral verification, observation, supervised conditions, or a combination?
\item What are the consequences for validity, reliability, fairness, accessibility, staff and student workload, privacy, and professional relevance?
\end{enumerate}

\begin{figure}[htbp]
\centering
\includegraphics[width=0.70\textwidth]{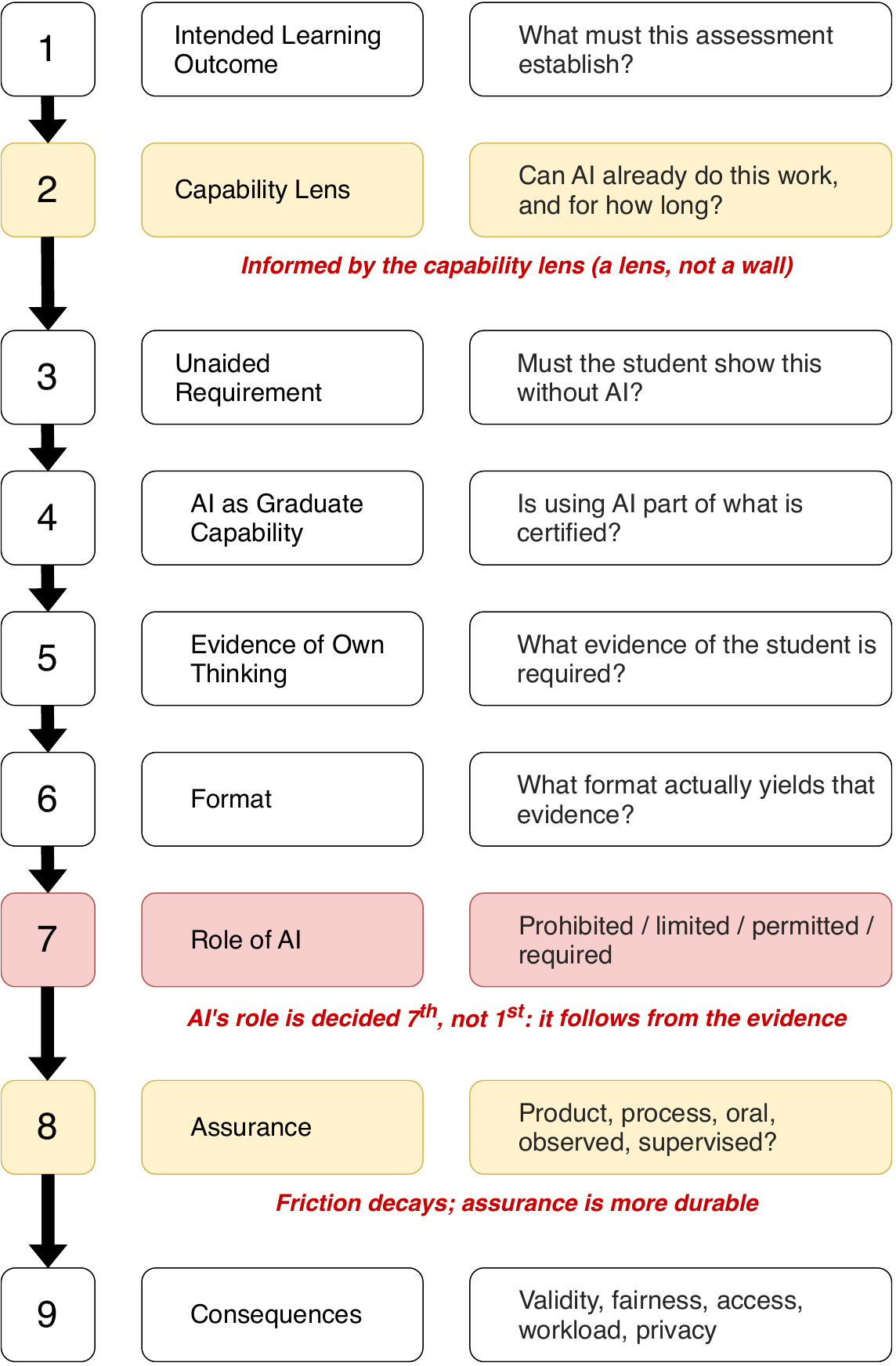}
\caption{The assessment-design decision chain. The role assigned to AI is the seventh decision, not the first: it follows from the evidence the outcome requires. Step~2 is informed by the capability lens; the chain's weight rests on the assurance decided at steps~8 and~9.}
\label{fig:chain}
\end{figure}

The order matters. The role assigned to AI is the seventh question here, not the first, because it should follow from the evidence the outcome requires. This follows \citet{klapproth2026redesigning}, but the chain places two further decisions after it: how learning is assured, and at what cost. Steps~8 and~9 force the designer to name the assurance mechanism and to count its costs, which is where the trade-offs of RQ4 enter the design rather than being discovered after it.

\subsection{Design moves, and the friction-versus-assurance distinction}

The chain yields two kinds of design move, and keeping them apart is essential. The first kind raises the cost or lowers the quality of an AI-produced answer: requiring the student's own collected or local data, requiring reflection on their own result or error, mandating very recent or in-class-only material, requiring the student to pose the problem rather than only answer it, or requiring a metacognitive account of their reasoning. These are friction, and friction decays as models and tools improve, so it should be treated as useful but temporary. This is a contested position. \citet{gonsalves2025contextual} argues that contextual design, built on practical, situational, experiential, interdisciplinary and ethical context, makes assessment more resilient because AI struggles with complexity and personal interpretation. That argument rests on capability evidence from 2023; the RQ3 evidence and capability coding suggest current tools now complete many such tasks. The disagreement is therefore about time: contextual tasks may resist some models, but this paper treats that resistance as temporary because it depends on what models cannot yet do. A no-AI rule backed only by friction is weak for a further reason: when a rule is likely to be broken, marks are interpreted on the false assumption that it was kept \citep{dawson2024validity}. The second kind depends far less on current AI limits: in-person supervised conditions, oral or viva verification with unpredictable, outcome-mapped questioning, and programme-level validity. These are assurance. They are more durable than friction, but not permanent: validity is a claim argued from evidence rather than a fixed property \citep{dawson2024validity}, and as assurance methods become standard, ways of defeating them can be expected to develop too \citep{elkhodr2026assurance}. Remote proctoring is not included here: there is little peer-reviewed evidence that it detects misconduct \citep{dawson2024validity}, and it controls the environment rather than evidencing a student's own reasoning \citep{elkhodr2026assurance}. Process and portfolio evidence sits between the two kinds. It makes the student's work visible and supports an oral or supervised check, but when it is produced in unsupervised take-home conditions it can be reconstructed after the event and does not on its own verify who did the reasoning \citep{elkhodr2026assurance}. The policy evidence shows institutions already building in such checks, with 21 of the 55 documents invoking oral or process verification, of which the oral and supervised component is the assurance and the process component supports it; the strongest designs, at the University of the Witwatersrand, ZHAW and the University of Sydney, pair an open task with a secured moment. The framework's recommendation is to build on assurance and to use friction only as a supplement.

\subsection{Worked example: the Bandura task}

The framework is made concrete through one assessment, deliberately chosen because it is the kind of task AI performs most easily: a request to describe Bandura's social learning theory in 500 words. Table~\ref{tab:bandura} traces five redesigns of this task. The progression is not from worse to better in the abstract, but from a design whose graded product is exactly what AI produces (Version~A) toward designs that either secure a moment of unaided reasoning or make the student's judgement about AI the object of assessment (Versions~C to~E). The example shows that the same topic can sit anywhere on Typology~2, and that the choice is driven by what evidence the outcome requires, which is the framework's central claim.

\begin{table}[htbp]
\centering
\footnotesize
\caption{The Bandura task redesigned across the framework. Each version answers the decision chain differently.}
\label{tab:bandura}
\begin{tabular}{@{}p{1.7cm} p{4.1cm} p{4.1cm} p{3.9cm}@{}}
\toprule
\textbf{Version} & \textbf{Redesign of the task} & \textbf{What it now assesses} & \textbf{How learning is assured} \\
\midrule
A. Conventional essay & ``Describe Bandura's theory in 500 words.'' & Recall and summary of public knowledge; squarely in the AI-strong zone & Not assured; the graded product is what AI produces \\
B. Evidence-based comparison & Using two specified studies, compare Bandura's theory with an alternative and judge which better explains a supplied case & Selection and weighing of evidence and argument against a specific case & Not assured; the supplied case is not in training data, which adds friction, but AI can still draft a plausible answer, so this is friction, not assurance \\
C. Process assessment & Submit research question, search strategy, annotated sources, plan, draft, final and reflection & The process of inquiry over time, not only the final text & Partly assured; staged artefacts make the process visible but can be reconstructed, so pair with a short oral or supervised check \\
D. AI-integrated & Given an AI-written account of Bandura, find its errors, verify claims against sources, improve it and justify the changes & The student's evaluation, verification and AI literacy & Partly assured; it shifts the graded object to the student's evaluation of AI, but as an unsupervised product (which AI can also critique) it needs a paired oral or supervised check \\
E. Hybrid assurance & Open written task followed by a short oral defence of specific choices made in the work & Product quality plus individual understanding of the submitted work & Assured by oral verification of authorship and understanding; AI can write the product but cannot sit the viva \\
\bottomrule
\end{tabular}
\end{table}

\subsection{Scope and limits of the framework}

The framework is a reasoned synthesis offered for testing, not a validated instrument. Three limits bound it. First, like most of the design literature in RQ3, it has not been empirically trialled for whether it improves assessment validity in authentic settings, so it is a hypothesis about good design rather than evidence of it. Second, its assurance mechanisms carry the costs catalogued in RQ4: oral and process verification are staff-intensive and can disadvantage non-native speakers and students needing accommodations, so applying the framework well is partly a resourcing and inclusion problem, not only a design one. Third, the capability lens that informs step~2 depends on a boundary that moves, so the framework must be re-run as models change rather than treated as settled. Used with these limits in view, it provides what the corpus otherwise lacks: a single, ordered way to decide what an assessment should establish and how to assure it, rather than a rule about whether students may use AI.

\section{Findings, RQ1: the problems for validity, integrity and equity}

\subsection{What the evidence base is}

Twenty-two of the 41 literature sources bear on RQ1. Of these, 8 are empirical studies, 9 are conceptual, 2 are commentary, 2 are reports and 1 is a review; 3 were coded as high-strength, 16 medium and 3 low. The high-strength sources are the two detection-testing studies \citep{weber2023testing,liang2023gpt} and the EU AI Act \citep{EUArtificialIntelligenceAct}; the Act is high-strength as an authoritative legal source, not as empirical evidence of prevalence or harm, and it is read here as the regulatory anchor rather than as a problem study. The problem framing is therefore supported by a small core of strong empirical work on detection, surrounded by a larger body of survey and conceptual material.

\subsection{Prevalence: AI use in assessment is now near-universal}

Student use of generative AI for assessed work has become near-universal within three years. The 2025 UK student survey \citep{StudentGenerativeAISurvey2025} reports that 92\% of students use AI in some form and 88\% have used it for assessments, up from 66\% and 53\% a year earlier. The 2026 UK survey \citep{StudentGenerativeAISurvey2026} puts near-universal use at 95\%, with 94\% using it for assessed work. The 2026 Digital Education Council global survey \citep{DigitalEducationCouncilAIinHigherEducati} reports 88\% of students and 77\% of faculty using AI. The FDUSP report \citep{USPFDUSP2024}, a small single-faculty survey of 94 law students, found about 65\% reporting AI use and 71\% saying they do not indicate in their work when they have used AI. These figures are self-reported and the largest surveys are UK-based, with the FDUSP data from Brazil and the Digital Education Council survey global, so they should be cited as indicative; but the direction is unambiguous, and it is the premise for treating presumed use as the realistic default rather than an edge case.

\subsection{Validity: the graded product no longer evidences the process}

The central problem is not cheating but validity \citep{dawson2024validity}. When a universally available tool can produce the graded artefact, the mark no longer certifies that the student performed the intended work. The point is demonstrated directly: \citet{cotton2024chatting} used ChatGPT to draft parts of an academic paper to show that LLMs produce plausible academic text, and \citet{dawson2020defending} had already identified reusable, unsupervised, one-right-answer tasks as the least secure. Evidence from higher education now points the same way. In a five-year natural experiment in one compulsory undergraduate course (N = 1066), replacing an AI-permissive take-home examination with a supervised in-person one, with content, outcomes, grading and examiner held constant, raised the failure rate from between 2 and 6 percent across 2021 to 2024 (2.1 to 5.7 percent) to 18.4 percent in 2025 \citep{brattli2026validity}, although the authors read this as association and note the format change coincided with other differences. In a randomised controlled trial (120 enrolled; 85 completed the delayed test), using ChatGPT as a study aid rather than traditional methods lowered retention on a surprise test 45 days later (57.5 versus 68.5 percent correct; $d = 0.68$) \citep{barcaui2025crutch}. A larger school-level trial gives the clearest causal picture: in a randomised trial of nearly 1,000 secondary-school mathematics students in T\"urkiye, unguided GPT-4 access raised practice performance by 48 percent but lowered scores on an unaided exam in the same session by about 17 percent against the control, whereas a guarded AI-tutor condition raised practice performance further without that drop \citep{bastani2025guardrails}. The validity threat these studies name, that measured performance can rise while unaided ability falls, is the one this paper treats as data. \citet{MOLLICKThefutureofeducationinaworldofAIb} frames the same point through the calculator analogy: the capability is permanent, so assessment must change rather than resist.

\subsection{Integrity: detection is unreliable, so the misconduct framing is fragile}

The integrity response that assumes detectable cheating rests on tools that do not work reliably. \citet{weber2023testing} tested 14 detection tools (12 free and two commercial) and found none reliable, with all below 80\% accuracy and false-positive rates ranging up to 50\%. \citet{liang2023gpt} found an average false-positive rate of 61.3\% for non-native English essays against near-perfect accuracy for native essays. Classroom accounts \citep{MOLLICKThefutureofeducationinaworldofAIb} show that light editing defeats detectors. The older contract-cheating literature compounds the problem: \citet{lancaster2008phenomena,lancaster2017rethinking} show that outsourced work is original and so evades text-matching, and that even invigilated examinations are not a complete safeguard. The implication for RQ1 is that framing the problem as detectable misconduct is fragile, because the detection it depends on is weak.

\subsection{Equity: detectors and access both fall unequally}

The problems are not neutral across students. On detection, the false-positive burden lands disproportionately on non-native English writers \citep{liang2023gpt}, which turns an unreliable tool into a discriminatory one. On access, several sources report that benefits and use skew by background: \citet{kasneci2023chatgpt} note multilingual bias and financial barriers; \citet{chan2023students} report student fears that AI widens the gap between rich and poor; the \citet{OECDDigitalEducationOutlook2026} finds benefits accruing to higher-performing and higher-socioeconomic students; and \citet{li2025integrating} report a stark adoption gap of about 47\% of institutions in high-income countries against 8\% in low-income countries. The UK surveys add a demographic dimension: women report worrying more than men about false accusations, and students from lower social grades report using some AI tools less \citep{StudentGenerativeAISurvey2025,StudentGenerativeAISurvey2026}. Restriction has its own equity cost, since blanket bans can block students who rely on assistive technology \citep{perkinsai}. The EU AI Act \citep{EUArtificialIntelligenceAct} places education in its high-risk category precisely because poorly designed systems can infringe the right to education and entrench discrimination.

\subsection{The three problems interact}

Validity, integrity and equity are not separable. Near-universal prevalence undermines a detection-first integrity response; unreliable detection then creates an equity harm through biased false positives; and unequal access makes an openly AI-permissive design unfair unless tools are provided. A policy that fixes one problem in isolation tends to worsen another, which is why the trade-offs in RQ4 follow directly from this section.

\subsection{Link to the propositions}

The RQ1 evidence supports the propositions most strongly where it concerns detection. The futility of detection (relevant to P1 and P2) and the case for testing unaided ability directly (P3) are both well supported by the strong detection studies and the prevalence data. The clearest neutral case is the EU AI Act, which regulates rather than endorses any particular response. The mixed cases, including \citet{dawson2020defending} and \citet{lancaster2017rethinking}, retain a role for secure assessment, which qualifies any claim that the problem is solved by redesign alone.

\section{Findings, RQ2: how universities and sector bodies are responding}
\label{sec:rq2}

\subsection{What the corpus is, and what it is not}

This section draws on 55 policy and guidance documents collected by purposive sampling for variation, not representativeness. The reason this matters: the counts below describe this sample, so they must not be read as population estimates for all universities. The corpus mixes document types: 28 are university guidance pages, 8 are formal university policies, 4 are sector or quality-assurance body documents, 3 are professional-body statements from publisher and editorial bodies (ICMJE, COPE and Springer Nature), and 12 are other grey literature, of which 7 are journal or publisher AI policies. The sample is also weighted toward the United Kingdom and the Anglophone world, so the geographic claims are illustrative rather than global. Two of the documents in this corpus are themselves systematic mappings of UK policy rather than single-institution statements: \citet{atkinson2024generative} analyse the generative-AI policies of all 24 Russell Group universities, and \citet{tao2026assessment} map 132 documents across 76 UK institutions. Because they sit inside the sample, they are not independent corroboration, but their larger-scale findings align with the patterns reported below. Both describe most institutions opting out of AI detectors and near-universal requirements to acknowledge AI use. \citet{tao2026assessment} additionally report convergence on graded permission categories, including traffic-light tiers and the AI Assessment Scale, and a shift toward process- and oral-based assessment, and characterise UK governance as assessment-centred and integrity-first.

\subsection{Outright prohibition is now a minority response}

Most documents permit generative AI use under conditions rather than banning it. Of the 55 documents, 24 take a default stance of "permit with limits" and 14 devolve the decision to individual staff or programmes; only 4 prohibit by default, and 4 position AI as expected or integral. Read through the paradigm coding (Typology 1), the same pattern appears: 20 documents combine several approaches, 17 centre on disclosure, 7 on redesign, 7 on integration, and only 2 sit squarely in the older "detect and punish" posture. The sector has largely moved past the question of whether students may use these tools and on to the question of under what conditions.

\subsection{Detection is avoided or unaddressed, and almost never mandated}

Among documents that take any position on AI-text detectors, the clear majority reject reliance on them. Thirty-two of the 55 documents are silent on detection. Of the 23 that do address it, 16 explicitly avoid or ban detectors, 5 treat detector output as an indicator only and never as sole evidence, 1 relies on lockdown or invigilation instead, and just 1 requires detector use. The institutions that explicitly reject detectors include the University of the Witwatersrand, which disabled the Turnitin AI detector across the institution in 2023, ZHAW, which prohibits AI-text detection software for assessment outright, Stellenbosch University, and Arts University Bournemouth. The single institution that requires detector use is United Arab Emirates University, whose policy obliges faculty to run AI-text detection tools and penalise unreflective reuse. This near-consensus against detection, with one clear outlier, is one of the strongest patterns in the corpus, and it aligns with the empirical literature on detector unreliability and bias against non-native writers \citep[for example][]{weber2023testing,liang2023gpt}.

\begin{figure}[htbp]
\centering
\includegraphics[width=\textwidth]{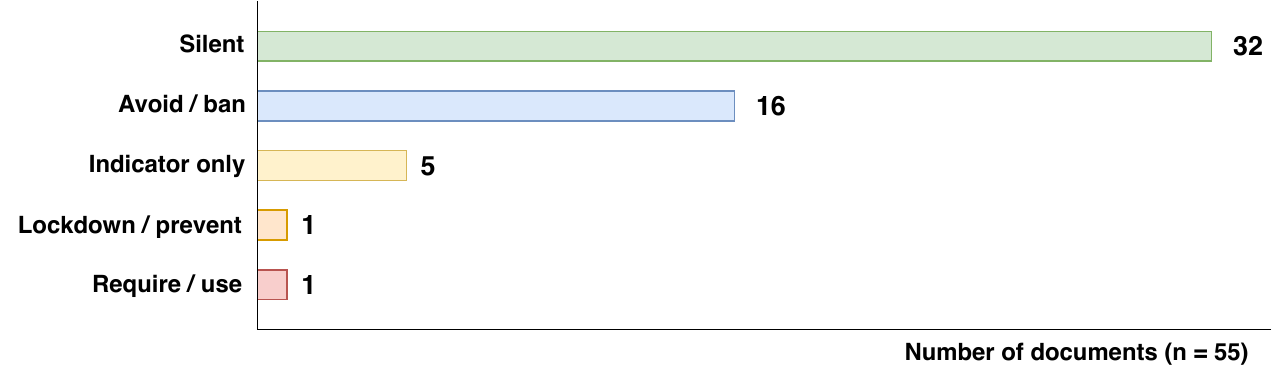}
\caption{Position on AI-text detection across the 55 policy documents. Thirty-two are silent; of the 23 that take a position, 16 avoid or ban detectors and only one (United Arab Emirates University) requires their use.}
\label{fig:detection}
\end{figure}

\subsection{Institutions are converging on structured permission scales}

Rather than a single allow-or-ban rule, a large share of documents define graded categories of permitted AI use. Thirty-one of the 55 documents record some form of defined category or tier, though the granularity varies widely. The clearest named schemes include University College London's three categories of use, the University of Wollongong College Hong Kong's three tiers of prohibited, somewhat permitted and expected use, United Arab Emirates University's prohibited, assistive and integral tiers, the four-level AI Assessment Scale used in the University of the Witwatersrand guidance, and ZHAW's nine-level glossary running from "AI-free" to "AI-produced." Harvard and Carnegie Mellon instead offer menus of example syllabus policies from which staff choose. The practical implication is that the field is standardising on scales of permitted involvement, which gives students and staff a shared vocabulary that a binary rule cannot.

\subsection{Disclosure is widely required but not standardised}

Most documents that permit AI use require students to declare or acknowledge it, yet they prescribe very different forms of declaration. Some require only a general statement, while others, such as Hong Kong Metropolitan University and ZHAW, require a listed declaration of each tool used and its purpose, and the University of Toronto and others ask for an appendix of prompts and outputs. This matters for the paper's second proposition, which holds that disclosure questions produce low-value information. The corpus does not settle that claim. The widespread adoption of disclosure shows institutions treat it as valuable, which cuts against the proposition; but the absence of any common standard, and survey evidence that students often do not comply with declaration requirements in practice (74 percent of students at one UK business school did not complete the mandatory AI-use declaration on their coursework coversheet \citep{gonsalves2025declarations}, and 71 percent of the 94 respondents in the small FDUSP law-faculty survey said they do not indicate in their work when they have used AI), suggests that declared disclosure and actual disclosure may diverge, which is consistent with the proposition's underlying worry.

\subsection{Redesign and oral or process verification recur as the main mechanism}

Where documents move beyond rules about tools, they converge on changing the assessment itself. The richest examples, from the University of the Witwatersrand, ZHAW, Stellenbosch and Arts University Bournemouth, recommend assessing the process rather than only the product, securing a small number of key moments for unaided ability, and using oral defence or interactive discussion to verify authorship. This is the clearest point of contact with the paper's first and third propositions, that redesign is more sustainable than detection and that unaided ability should be tested directly in controlled conditions. The corpus supports both as the direction of institutional travel, while showing that few institutions have yet operationalised them at scale.

\subsection{Variation by country and type, and its limits}

The sample shows real variation. The United Arab Emirates University detector mandate contrasts sharply with the United Kingdom and South African avoidance of detectors; Swedish and Irish guidance ties any restriction to intended learning outcomes; several North American pages devolve almost entirely to the instructor. By institution type, the clearest pattern is prohibition: all four default prohibitions in the corpus come from research-intensive universities (Toronto Mississauga, one Imperial College faculty, Oxford and Stanford), while none of the 5 teaching-intensive institutions or 18 non-university bodies prohibits by default. Research-intensive institutions are also the most likely to devolve the decision to individual staff (12 of 32), whereas non-university bodies, mainly publishers and sector agencies, overwhelmingly permit with limits (11 of 18). The non-research cells are small, so these are leads rather than conclusions, but prohibition is concentrated where high-stakes, discipline-specific assessment is most common. Two cautions apply. First, the sample is UK-weighted, so apparent national patterns outside the UK rest on few cases and should be stated as leads, not conclusions. Second, every document in this corpus is declared policy, not observed practice, so the section describes what institutions say they do, and the gap between policy text and classroom reality remains an explicit limitation.

\subsection{Link to the propositions}

Taken together, the corpus leans toward the post's propositions without confirming them. Coded by direction, 36 of 55 documents support the propositions, 14 are mixed, 2 challenge them and 1 is neutral. The support is strongest for redesign over detection (P1) and for securing unaided ability directly (P3). It is weakest, and most contested, for the claim that disclosure is low-value (P2), where institutional behaviour and the post pull in opposite directions. Because the coding was framed around the propositions, the high support count should be read as a prompt to foreground the mixed and challenging cases rather than as a tally in the propositions' favour.

\section{Findings, RQ3: assessment-design models for a world of presumed AI use}

\subsection{The design evidence is mostly proposal, not proof}

Read the design literature for what kind of evidence it is before reading it for what it recommends. Of the 27 RQ3 sources, 15 are conceptual or theoretical, 5 are empirical studies, 3 are reviews, 3 are commentary or editorial, and 1 is a sector report. Only 2 were coded as high-strength evidence \citep{MOORHOUSE2023ASSESS,JIAO2024LANDSCAPE}, with 23 at medium and 2 low. The practical consequence: this section describes a field rich in frameworks and arguments but thin in validation. Most proposed designs have not yet been tested for whether they actually secure assessment validity in authentic settings, so they are best read as reasoned proposals rather than proven solutions.

\subsection{AI-era design rests on pre-AI assessment theory}

The strongest claim the corpus supports is one of continuity, not novelty. The design models proposed for generative AI re-apply principles that long predate it. Authentic assessment built from real-world tasks \citep{wiggins1990case}, constructive alignment of tasks to intended outcomes \citep{biggs1996enhancing}, sustainable assessment and the development of students' evaluative judgement \citep{boud2006aligning}, programmatic assessment that combines many information sources with human judgement \citep{schuwirth2011programmatic}, and designing assignments so that cheating is engineered out \citep{lancaster2008phenomena,bretag2016handbook} all reappear in the AI-era guidance. The argument of \citet{dawson2020defending}, that assessment design should take priority over assessment security, is the hinge between the two literatures, and it is echoed across the newer sources. The implication for the paper is that the question is less "what new assessment does AI require" and more "which established principles now carry more weight." \citet{klapproth2026redesigning} makes the same argument for constructive alignment specifically: the framework still holds, but outcomes must now state whether the performance is with or without AI. As \citet{ransome2026reveals} puts it, generative AI ``hasn't created assessment problems''; it has exposed them.

\subsection{The field is converging on graded scales and two-lane designs}

Both the literature and the policies cluster on two related structures. The first is a graded scale of permitted AI involvement, exemplified by the AI Assessment Scale \citep{perkinsai}, which runs from no AI to full AI integration, and mirrored in the graded institutional schemes catalogued in Section~\ref{sec:rq2}. The second is a two-lane model that separates secure, supervised assessment from open, AI-permitted assessment at programme level, introduced as the two-lane approach \citep{liubridgeman2023twolane} and formalised in the Sydney Assessment Framework \citep{bridgemanliu2025framework}, and argued more formally by \citet{corbin2025talk}, who locate validity in the structure of an assessment programme rather than in individual tasks. The policy coding reflects this convergence: 31 of 55 documents promote a mixed assessment logic and a further 7 a hybrid logic, against only a handful promoting a single secure or open model. Scales and lanes, not single rules, are the emerging design vocabulary. The two-lane model has a direct critic. \citet{curtis2025twolane} treats it as an all-or-none choice between banning AI and permitting any use, and argues that a third lane, permitting AI within limits, is essential; he keeps both the secure and the open lane and ties the middle lane to the AI Assessment Scale \citep{perkinsai}. The graded scales in the corpus supply such a middle lane. His case for it, however, rests on enforcing limits through layered detection, including imperfect AI detectors used as one source of evidence among several, and he explicitly rejects the claim, made by \citet{dawson2024validity}, that assessments which cannot prevent AI use have little value for high-stakes purposes. This paper takes a different view of enforcement (Sections~3.5 and~5.3): a limited-AI rule on unsupervised work is friction unless it is paired with an assurance check.

\subsection{Oral verification and supervised checks are the assurance mechanism; process evidence supports them}

Where designs move from permission to assurance, they rely on seeing the process and testing the person directly. Twenty-one of the 55 policies mention oral or process verification. The most developed examples include King's College London, which uses investigatory vivas and academic-integrity meetings to probe authorship; University College London's investigatory viva route; the University of Sydney's interactive oral conversations within secure assessment; and the detailed process-and-oral provisions at the University of the Witwatersrand and ZHAW, both of which recommend replacing fully AI-producible tasks with oral defence and staged process assessment. Sector bodies point the same way, with TEQSA, the 2025 and 2026 EDUCAUSE Horizon Reports \citep{robert20252025,robert2026horizon} and the OECD \citep{OECDDigitalEducationOutlook2026} review all foregrounding process-focused and evidence-centred assessment. In the literature this is the modern face of programmatic assessment: validity is assured at a few secure or synchronous moments and through evidence of the student's own working, rather than at every task.

\subsection{Few sources claim a durable AI-proof task}

The corpus is honest about the limits of designing around current model weaknesses. Of the 27 RQ3 sources, only 1 treats its target work as sitting in a zone where AI is weaker, 4 place it in a zone AI already handles well, and 14 are mixed, with the remainder not applicable. In plain terms, almost none argue that a task type is robustly beyond AI; most accept that capability is moving. Several authors say so directly: \citet{dawson2020defending}, Carnegie Mellon's guidance, and \citet{MOORHOUSE2023ASSESS} all warn that designing assignments around what AI cannot do today is a temporary fix, not a sustainable one. This connects to the paper's capability-lens argument (Section~\ref{sec:framework}): the durable basis for design is assurance of the individual's reasoning, not a search for tasks the model will fail.

\subsection{Authentic assessment alone is not sufficient}

A specific and important finding is that authentic assessment, long treated as the answer to misconduct, does not by itself block AI. A small experiment at two UK universities found that markers could not reliably distinguish work with GenAI input from human-authored work, with detection accuracy ranging from 33 to 86 percent, and that the level of authenticity made no difference \citep{kofinas2025impact}. The review evidence \citep{MOORHOUSE2023ASSESS,JIAO2024LANDSCAPE} and the capability coding point the same way: many authentic, take-home and higher-order tasks can be completed by current tools. The design response in the strongest sources is therefore not "make it authentic" but "make it authentic and assured," pairing an authentic open task with a secure or oral component. This correction matters, because it overturns a common assumption in the earlier integrity literature.

\subsection{Integration is a separate strand, not the same as redesign}

A distinct group of sources treats AI not as a threat to design around but as content to assess. \citet{kasneci2023chatgpt}, \citet{chan2023comprehensive}, and the 2026 sector reviews \citep{DigitalEducationCouncilAIinHigherEducati,OECDDigitalEducationOutlook2026} describe assessments that require students to use, critique, verify and improve AI output, and that assess AI literacy itself. This "integrate" strand overlaps with redesign but answers a different question: not how to preserve unaided ability, but how to assess the new capability of working well with AI. Keeping the two strands separate in the paper avoids conflating "redesign so AI cannot do it" with "redesign so AI is the thing being assessed."

\subsection{The evidence gap and the state of adoption}

Two limitations run through the corpus. First, adoption lags the frameworks. The Digital Education Council's 2026 global survey \citep{DigitalEducationCouncilAIinHigherEducati} and the 2026 student survey \citep{StudentGenerativeAISurvey2026} both report that AI-resilient and AI-integrated redesign is under way but far from universal, and that students lack clarity on what is permitted. Second, several of the most specific frameworks are grey literature or not yet empirically validated. The PROTECT framework \citep{PROTECTassessmentframework} is a teaching-resource page, and the four-pillar framework of \citet{nikolic2026designing}, though peer-reviewed, states that its own claims have not been empirically tested. The paper's own design framework (Section~\ref{sec:framework}) sits squarely in this strand and should be positioned as a synthesis of these proposals, offered for testing rather than as settled practice.

\subsection{Link to the propositions}

The design literature leans toward the post's propositions. Of the 27 RQ3 sources, 18 support them, 7 are mixed, 1 is neutral and 1 not applicable. Support is strongest for redesigning assessment where AI can do the work (P1) and for testing unaided ability directly in secure conditions (P3). The mixed cases are the ones to foreground: \citet{dawson2020defending}, \citet{lancaster2017rethinking} and \citet{cotton2024chatting} all retain a role for secure examinations or even detection alongside redesign, which qualifies any simple reading that redesign alone is sufficient.

\section{Findings, RQ4: trade-offs}

\subsection{What the evidence base is}

Twenty-one of the 41 literature sources bear on RQ4, alongside the equity and detection fields of the policy corpus. The trade-offs below are drawn from these sources. They are reported as tensions that any assessment design must balance, not as problems with a single correct resolution, because the evidence shows that improving one dimension usually costs another.

\subsection{Learning versus performance: the offloading trade-off}

Permitting AI can raise output while lowering learning. The same secondary-school trial \citep{bastani2025guardrails}, in which unguided AI access produced large gains in practice performance but lower scores on an unaided exam in the same session, is cognitive offloading measured directly. The trade-off is therefore real and not rhetorical: the same tool that makes a student more productive now can make them less capable later. \citet{MOLLICKThefutureofeducationinaworldofAIb} argues the opposite reading, that managed use builds durable skill, which is why the resolution depends on design rather than on allowing or banning the tool outright.

\subsection{Integrity versus fairness: the detection dilemma}

Detection can deter some misconduct, but at a fairness cost that the strongest evidence shows is severe. Because detectors misfire disproportionately on non-native writers \citep{liang2023gpt}, and detection tools are unreliable more generally \citep{weber2023testing,perkinsai}, buying integrity through detection spends equity. Even an accurate detector carries a validity cost: by flagging some submissions it can give assessors false confidence that unflagged students did not use AI \citep{dawson2024validity}. \citet{nikolic2026designing} name a sharper version, an equity paradox in which moving from detection to oral defence reduces the false-positive harm to non-native writers but can introduce new biases in oral performance. There is no free move here: each verification method shifts the burden onto a different group of students. One partial answer is to aim for equivalent assurance rather than identical format: a student who cannot fairly demonstrate a capability orally may do so through a supervised written scenario or a live walkthrough that tests the same reasoning \citep{elkhodr2026assurance}.

\subsection{Access versus restriction}

Both permitting and restricting AI carry an access cost. Permissive designs advantage students who can afford better tools and already know how to use them, which widens existing gaps \citep{kasneci2023chatgpt,chan2023students,DigitalEducationCouncilAIinHigherEducati}. Restrictive designs can block students who depend on assistive technology, since a blanket ban removes tools some learners need \citep{perkinsai}. The resolution that appears in the policy corpus is to provide approved tools at no cost, as Uppsala University requires whenever AI use is expected, which converts an access problem into an institutional cost.

\subsection{Assurance versus workload and inclusion}

The verification methods that best assure validity are the hardest to scale and can themselves exclude. Oral and process-based verification, recommended by the University of the Witwatersrand and ZHAW and used at King's College London and the University of Sydney, is time-intensive and difficult at large class sizes. An oral component is not assurance by itself. Open question time after a presentation gives inconsistent evidence unless questions are planned, mapped to outcomes and applied consistently \citep{elkhodr2026assurance}, and swapping written work for orals without changing what is assessed can reproduce the same problems in a new format \citep{ransome2026reveals}. Moving assessment in-class or oral can also disadvantage non-native speakers and students needing accommodations, as the Carnegie Mellon and \citet{MOORHOUSE2023ASSESS} material warns; an oral check that some students cannot fairly complete lowers validity for those students \citep{dawson2024validity}. Assurance is therefore not free: it is bought with staff time and, unless designed carefully, with reduced inclusion.

\subsection{Privacy and data protection}

Several mechanisms create data-protection exposure. Running student work through AI detectors, or requiring students to use third-party AI tools, transmits their work and sometimes personal data to external systems. Institutions respond by restricting processing to approved, reviewed systems, as KU Leuven and ZHAW do, and the FDUSP report \citep{USPFDUSP2024} treats privacy and data protection as a primary governance concern. The trade-off is between the pedagogical value of a tool and the data risk of using it, which institutional procurement rather than individual staff are best placed to manage.

\subsection{Authenticity versus securability}

Authentic, take-home tasks that resemble real-world work are harder to secure than narrow invigilated tasks, and narrow secure tasks are often less valid as evidence of real capability. \citet{dawson2020defending} frames this directly as design versus security, and the design literature resolves it not by choosing one but by pairing an authentic open task with a small number of secure or oral checkpoints. The cost of that resolution is complexity: a programme must now manage two kinds of assessment rather than one.

\subsection{Innovation versus regulation and sustainability}

Two further trade-offs sit at the edge of the assessment question but belong in a full account. Regulation constrains innovation: the EU AI Act places education in its high-risk category, imposing transparency, human-oversight and documentation duties that raise the cost of deploying AI in assessment \citep{TheEUAIActandItsContractualandEducationa}. Sustainability constrains use: generative AI carries an environmental cost that institutions such as KU Leuven ask staff to weigh before use. Both are minor strands in the current corpus, not central findings.

\subsection{Link to the propositions}

The trade-offs support the paper's underlying move from capability-chasing to assurance. Because every verification method and every permission setting shifts a cost onto some group, no single lever resolves the problem, which is the practical case for assurance-based design over the search for an AI-proof task. The trade-offs also qualify the propositions: the detection dilemma and the assurance-workload tension show why some institutions retain secure examinations alongside redesign, which is a mixed rather than a wholesale endorsement of the post's position.

\section{Findings, RQ5: future directions}

\subsection{Why this section is written as scenarios, not predictions}

The forward-looking evidence is the weakest in the corpus, so this section offers scenarios rather than forecasts. Of the 18 sources coded to RQ5, 11 are conceptual, 3 are empirical, 2 are commentary, 1 is a review and 1 is a report, and only 1 is high-strength \citep{weber2023testing}, with 16 at medium strength and 1 low. The policy corpus reinforces the caution: 32 of the 55 documents carry a volatility note, being interim guidance, consultation drafts, or pages explicitly marked as under review, including the Quality Assurance Agency's state-of-the-nation work for 2026 to 2027 and several King's College London pages that label themselves guidance subject to rapid change. A field whose own documents say they are provisional cannot responsibly be predicted. The honest output is a small set of scenarios, each with the conditions that would make it more or less likely, and signposts to watch.

\subsection{Cross-cutting drivers}

Five forces shape every scenario below. First, model capability keeps moving, so any design that depends on a current weakness has a short half-life (see the capability lens in the conceptual framework). Second, regulation is arriving: the EU AI Act places education in its high-risk category, imposing transparency, human-oversight and liability duties \citep{TheEUAIActandItsContractualandEducationa}. Third, detection remains unreliable and is widely rejected, so it is unlikely to become the stabilising mechanism. Fourth, cost and access shape equity, because any design that assumes paid tools or heavy staff time redistributes advantage. Fifth, staff workload and scalability constrain what can actually be run at a large class size. The scenarios differ mainly in how these five forces resolve.

\subsection{Scenario A: bifurcation becomes the default}

In this scenario the two-lane model becomes the sector norm. Programmes separate a small secure lane that tests unaided ability under controlled conditions from an open lane in which AI use is permitted and often expected, with validity judged at programme level. The corpus already points this way through the Sydney framework, ZHAW's assessment rules, and the graduated AI Assessment Scale \citep{perkinsai}. This becomes more likely if two-lane designs prove scalable and if sector bodies endorse them in their forthcoming frameworks. It becomes less likely if the secure lane proves too costly or if it disadvantages non-native speakers and students needing accommodations, which would make the "secure" half the weak point. A strict two-lane version is also contested on design grounds: \citet{curtis2025twolane} argues an all-or-none split is insupportable and that a middle lane permitting limited AI is essential, and if that view prevails, bifurcation is more likely to arrive as a graded scale with a secured moment than as two pure lanes.

\subsection{Scenario B: AI-integrated assessment goes mainstream}

Here the dominant design assesses working well with AI rather than working without it. Students are required to use, critique and improve AI output, and AI literacy becomes an explicit outcome. The Digital Education Council and Pearson review \citep{TheNextEraofAssessmentAGlobalReviewofAIi}, built on 101 case studies, already identifies this as a leading direction, and \citet{kasneci2023chatgpt} and \citet{chan2023students} support it. It becomes more likely if the employability argument dominates and if institution-provided tools stabilise. It becomes less likely if the learning-versus-performance evidence hardens: a school trial found that unguided AI access raised practice scores but lowered scores on an unaided exam in the same session \citep{bastani2025guardrails}, and a higher-education trial found that unrestricted ChatGPT use as a study aid lowered retention 45 days later \citep{barcaui2025crutch}; if that pattern holds broadly, requiring AI use could be seen as harming the learning it is meant to serve.

\subsection{Scenario C: a return to secure and proctored assessment}

In this scenario institutions respond to a loss of trust by reinstating invigilated examinations, oral examinations and lockdown conditions as the centre of gravity, not the periphery. Because detection is discredited, the restoration takes the form of controlled conditions rather than detectors. A high-profile integrity crisis, a regulator demand for assured identity, or simple exhaustion with redesign could trigger it. It becomes less likely where the known costs of narrow secure assessment dominate: reduced validity, test anxiety, and exclusion of students who perform poorly in timed or oral settings, a cost flagged by \citet{dawson2020defending} and by the Carnegie Mellon and \citet{MOORHOUSE2023ASSESS} guidance. This scenario would partly vindicate the paper's third proposition, that unaided ability should be tested directly, while rejecting the first.

\subsection{Scenario D: programmatic and continuous validity}

Here validity is assured across a whole programme rather than at single tasks, drawing on programmatic assessment \citep{schuwirth2011programmatic} and the process-and-evidence designs now proposed \citep{OECDDigitalEducationOutlook2026,nikolic2026designing}. This scenario follows from treating validity as a claim built from chains of evidence across multiple tasks, occasions and assessors rather than secured by any single task \citep{dawson2024validity}. \citet{elkhodr2026assurance} describe a unit-level version, in which each outcome is verified at a point proportionate to its stakes rather than in a single final examination. Many low-stakes touchpoints, process evidence and periodic secure checkpoints together support a defensible judgement of competence. It becomes more likely if institutions invest in the assessment infrastructure and governance this requires, and if sector bodies promote programme-level rather than task-level validity. It becomes less likely if the data demands raise privacy and surveillance concerns, or if the staff effort proves unaffordable at scale.

\subsection{Scenario E: status-quo drift}

The null scenario is that little coordinates. Institutions continue with a patchwork of ad hoc rules, intermittent detector use and inconsistent disclosure requirements, and students remain unsure what is allowed. The 2026 Digital Education Council survey \citep{DigitalEducationCouncilAIinHigherEducati} is the warning sign: redesign is under way but far from universal, and a majority of students report that guidance is inconsistent across their assessments. This is the default if nothing changes, and it is the least stable, because it combines the equity harm of unreliable detection with the confusion of unclear rules. It would, in effect, vindicate the post's critique that detection and disclosure-policing produce low-value information without resolving the underlying validity problem.

\subsection{Signposts to watch}

Several observable developments would tip the balance between scenarios, and a review such as this should name them for future work. The first is whether sector bodies publish settled frameworks, with the QAA state-of-the-nation work and TEQSA's reform programme the ones to track. The second is any genuine change in detector reliability, in either direction, since a reliable detector would revive Scenario C through detection and a definitive rejection would entrench redesign. The third is enforcement of the EU AI Act in education, which would push institutions toward documented, auditable designs. The fourth is accumulating evidence on learning loss, which would constrain Scenario B. The fifth is the annual adoption and clarity data from large student surveys, which measures how far drift is giving way to coherence.

\subsection{Link to the propositions}

The scenarios that most fully realise the paper's propositions are bifurcation, programmatic validity, and AI-integrated assessment paired with secure moments, because each redesigns around presumed AI use (P1) while retaining direct tests of unaided ability where validity requires them (P3). The restoration scenario would support P3 while rejecting P1, and status-quo drift would vindicate the propositions' critique of detection and disclosure (P1 and P2) by negative example. No scenario supports a detection-first future, which is the clearest forward-looking conclusion the corpus allows. Because the RQ5 sources skew toward authors who advocate redesign, this section should be read with that selection bias in mind: the literature's confidence in a redesign future may run ahead of what institutions, constrained by cost and workload, will actually deliver.

\section{Discussion}

\subsection{Returning to the propositions}

The paper set out to test four propositions drawn from the anchoring post, rather than to endorse them. Across the two corpora the evidence supports the propositions unevenly, and the pattern of that unevenness is itself the main finding. In the policy corpus, 36 of 55 documents were coded as supporting the propositions, 14 as mixed and 2 as challenging; in the literature, 22 of 41 supported, 8 were mixed and 2 neutral. These counts come from coding judgements and should be read as the shape of the evidence, not as a measurement, but the shape is consistent: the propositions are strongest where they concern detection and redesign, and weakest where they concern disclosure.

\begin{figure}[htbp]
\centering
\includegraphics[width=\textwidth]{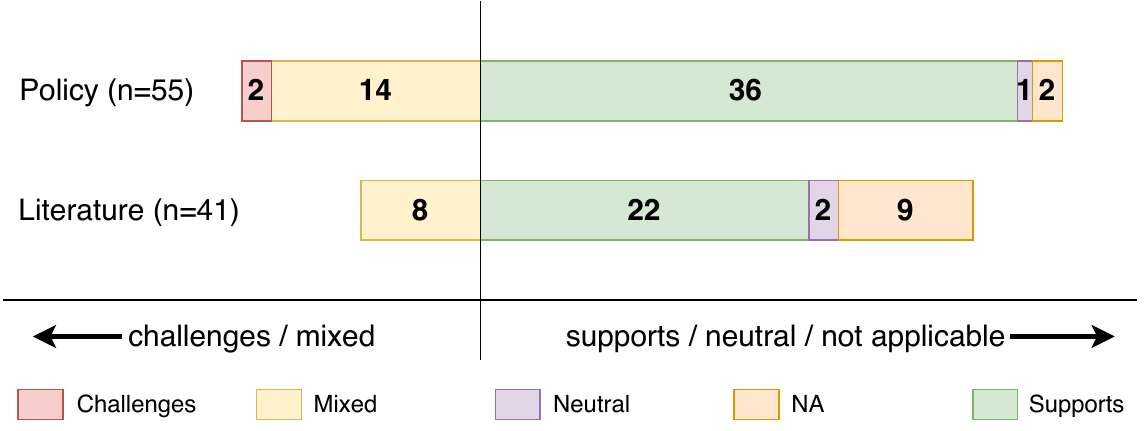}
\caption{Direction of coded support for the four propositions, by corpus. Bars show document counts; note the large ``not applicable'' share in the literature. The figure shows the shape of the evidence, not a measurement, and the counts rest on single-coder judgements.}
\label{fig:propdir}
\end{figure}

\textbf{P1, that redesign beats detection or banning where AI can do the work, is well supported but not unconditioned.} Institutional practice has moved decisively away from prohibition: only 4 of 55 documents prohibit AI by default, and among those taking any position on detection, 16 of 23 reject detectors while just one requires their use. The design literature points the same way. Yet the support is conditional. Several cautious sources, including \citet{dawson2020defending} and the two high-strength reviews, keep a role for secure assessment alongside redesign, and the capability evidence shows that redesign alone does not place a task beyond AI. The defensible version of P1 is therefore that redesign plus assurance, not redesign alone, is the sustainable response.

\textbf{P2, that disclosure questions produce low-value information, is the proposition the evidence most clearly complicates.} Institutions behave as though disclosure is valuable: most permissive policies require it, and several, including Hong Kong Metropolitan University and ZHAW, mandate detailed declarations. This cuts against P2. At the same time, there is no common standard for what disclosure should capture, and the survey evidence available points to low compliance in practice: 74 percent of students at one UK business school did not complete the mandatory AI-use declaration on their coursework coversheet \citep{gonsalves2025declarations}, and 71 percent of the 94 respondents to the single-faculty FDUSP study said they do not indicate in their work when they have used AI. The honest reading is that declared disclosure and actual disclosure diverge: disclosure may be valuable in principle while producing low-value information in practice, which is a more precise claim than the proposition as stated.

\textbf{P3, that unaided ability should be tested directly in controlled conditions, is supported as a component rather than as a whole system.} The emergence of two-lane and graduated designs, and the 21 of 55 policies that invoke oral or process verification, show institutions securing or evidencing specific moments of individual performance rather than securing everything. This is P3 realised as one lane within a programme, not as a return to invigilation across the board, and it is accompanied by real costs in workload and inclusion that the proposition does not acknowledge.

\textbf{P4, that the same logic applies to peer review, is the least evidenced.} The corpus contains publisher and editorial policies, including the ICMJE and COPE guidance that addresses AI use in peer review directly, and these treat AI use through disclosure and accountability rather than detection, which is broadly consistent with P4. Even so, the corpus holds too few peer-review-specific sources to test the proposition properly, so P4 is weakly supported rather than fully tested.

\subsection{What the paper concludes}

Three conclusions follow. First, the sector's centre of gravity has shifted from provenance to permission and design: the dominant question in institutional policy is now under what conditions AI may be used, not whether it was used, and detection has largely failed to become the stabilising mechanism. Second, the field is converging on a shared design vocabulary, built from graded permission scales and two-lane structures, with assurance provided mainly by oral verification and supervised checks, supported by process evidence; 31 of the 55 documents, more than half, already define named categories or tiers. Third, and cutting across both, the durable basis for valid assessment is the assurance of a student's own reasoning rather than the search for tasks that AI cannot perform, because the capability boundary moves and authentic tasks alone do not secure validity. The conceptual framework offered in this paper is an attempt to operationalise that third conclusion.

\subsection{Implications for practice and policy}

For institutions, the practical implication is to decide assurance before permission, building verification into the assessment sequence rather than appending it afterwards \citep{elkhodr2026assurance}. The framework's ordering, in which the role assigned to AI follows from the evidence an outcome requires, matches recent outcome-first guidance \citep{klapproth2026redesigning} and adds explicit decisions on assurance and cost. Many institutional policies in the corpus instead begin from a rule about permitted AI use; whether outcome-first design produces more valid assessment remains to be tested. For sector bodies, the implication is that coordination on shared scales and on the status of detection would reduce the student confusion that the 2026 survey evidence documents. For both, the trade-offs are not incidental: assurance costs staff time and can exclude, unequal access makes permissive designs unfair without provided tools, and detection spends equity for a weak integrity gain. A policy that names its assurance mechanism and budgets for its costs is more honest than one that bans or permits in the abstract.

\subsection{Limitations of the evidence and the reading}

The conclusions are bounded by how the study was done. The review is pragmatic and exploratory, not systematic: sources were gathered by targeted searching, citation chasing and purposive selection, so the corpora are not exhaustive and may carry selection effects, including an over-representation of authors who advocate redesign. Because collection was pragmatic, several closely related works were identified late and are discussed but not coded. The policy sample is weighted toward the United Kingdom and the Anglophone world, so cross-national claims are illustrative. All typology, capability and proposition codings were proposed by the extraction model and reviewed by a single author, so the counts await independent second-coding and should be read as indicative. The policy documents record declared policy, not observed practice, so the paper cannot speak to what staff and students actually do. And the design framework is a reasoned synthesis that has not been empirically trialled.
\subsection{Directions for research}

The gaps point to a research agenda. The most valuable next step is empirical: trials that test whether the design moves this paper synthesises, two-lane structures, process evidence and oral verification, actually improve assessment validity and for which students, rather than assuming they do. Such trials could build on validation programmes already proposed for assurance checkpoints \citep{elkhodr2026assurance} and for outcome-derived AI rules \citep{klapproth2026redesigning}. A second need is evidence on practice rather than policy, closing the gap between what institutions declare and what classrooms do. A third is a deliberately non-Anglophone and Global South sample, to test whether the patterns reported here hold where access, language and resource conditions differ. A fourth is direct study of AI in peer review, which this paper could only touch. Pursued together, these would convert the framework from a reasoned proposal into a tested one, which is the condition on which its value finally depends.

\section{Conclusion}

Generative AI has made the question that most assessment procedures are built to answer, whether a human produced a given piece of work, both unanswerable and beside the point. Building on the argument that validity matters more than cheating \citep{dawson2024validity}, this paper has argued that the problem worth solving is validity, not provenance. It has used an opinion piece that states a strong version of that case as a foil, restating it as four propositions and testing them against two corpora: 41 research and sector sources and 55 policy and guidance documents.

Three findings stand out. The sector has moved decisively away from prohibition and detection: only 4 of 55 documents prohibit AI by default, and among those taking a position on detection, 16 of 23 reject detectors while just one requires them. The field is converging on a shared design vocabulary of graded permission scales and two-lane structures, with assurance supplied mainly by oral verification and supervised checks, supported by process evidence, and 31 of the 55 documents, more than half, already define named categories or tiers. And the durable basis for valid assessment is the assurance of a student's own reasoning rather than the search for tasks that AI cannot perform, because the capability boundary moves and authentic tasks alone do not secure validity.

The paper's contribution is to test a strongly stated position against two kinds of evidence, and to turn the result into a design logic. Building on outcome-first guidance \citep{klapproth2026redesigning}, the conceptual framework orders the decision so that the role assigned to AI follows from the evidence an outcome requires, and then requires the designer to name the assurance mechanism and count its costs. The Bandura worked example shows the same topic redesigned from a task AI can simply complete to tasks that secure a moment of unaided reasoning or make the student's judgement about AI the object of assessment. Tested against this evidence, the propositions hold unevenly: redesign over detection is well supported but only as redesign plus assurance; testing unaided ability directly is supported as one lane within a programme; the claim that disclosure is low-value is complicated, since institutions treat disclosure as valuable while the survey evidence suggests students often do not complete the required declaration in practice \citep{gonsalves2025declarations}; and the extension to peer review is plausible but under-examined.

These conclusions are bounded by how the study was done. It is a pragmatic review of a purposive, UK-weighted, almost entirely English-language sample; its codings are single-coder judgements that await second-coding; its extraction was AI-assisted; and it describes declared policy, not practice. The framework it offers is a reasoned synthesis, not a validated instrument. The value of that framework now depends on the research it invites: empirical trials of whether two-lane designs, process evidence and oral verification actually improve validity and for which students; evidence on practice rather than policy; a deliberately non-Anglophone and Global South sample; and direct study of AI in peer review. Until that work is done, the honest claim is modest but useful: higher education should stop asking whether students used AI, and start deciding, outcome by outcome, what each assessment must establish and how it will be assured.

\section*{Acknowledgements}
The author thanks Dr Sylvia Terbeck (Liverpool John Moores University), whose public post of 20 September 2026 prompted this paper and serves as its anchoring position. The restatement of that position as four propositions, and the analysis, framework and conclusions that test them, including any that qualify the position, are the author's own.

\appendix
\section{Comparative policy matrix}
\begin{landscape}
\scriptsize
\setlength{\tabcolsep}{3pt}
\renewcommand{\arraystretch}{1.1}
\begin{longtable}{@{}>{\raggedright\arraybackslash}p{3.5cm} >{\raggedright\arraybackslash}p{1.2cm} >{\raggedright\arraybackslash}p{1.6cm} >{\raggedright\arraybackslash}p{1.45cm} >{\raggedright\arraybackslash}p{1.45cm} >{\raggedright\arraybackslash}p{1.3cm} >{\raggedright\arraybackslash}p{1.35cm} c >{\raggedright\arraybackslash}p{1.2cm}@{}}
\caption{Comparative policy matrix: 55 documents by stance, paradigm, assessment logic, detection position, defined categories (Cat.), and proposition direction. All cell text is shown in full. The Type column gives institution type for universities and, for non-university sources, the document type (sector body, professional body, publisher, research paper or grey literature). These reconcile with the source-type composition in Table~\ref{tab:corpus}, with one exception: the MIT committee report is a research-intensive university document but is classified as grey literature by source type, so Table~\ref{tab:corpus} counts it among the 19 non-university documents while it appears here among the 37 university-affiliated rows.}\label{tab:matrix}\\
\toprule
\textbf{Institution} & \textbf{Juris.} & \textbf{Type} & \textbf{Stance} & \textbf{Paradigm} & \textbf{Logic} & \textbf{Detect.} & \textbf{Cat.} & \textbf{Prop.} \\
\midrule
\endfirsthead
\multicolumn{9}{@{}l}{\footnotesize\itshape Table \ref{tab:matrix} continued}\\
\toprule
\textbf{Institution} & \textbf{Juris.} & \textbf{Type} & \textbf{Stance} & \textbf{Paradigm} & \textbf{Logic} & \textbf{Detect.} & \textbf{Cat.} & \textbf{Prop.} \\
\midrule
\endhead
\midrule \multicolumn{9}{r@{}}{\footnotesize continued on next page}\\
\endfoot
\bottomrule
\endlastfoot
TEQSA (Tertiary Education Quality and Standards Agency) & Australia & NA (sector body) & Integrate/\allowbreak{}AI-expected & Redesign & Multiple & Avoid/\allowbreak{}ban & N & Supports \\
University of Melbourne & Australia & Research-intensive & Permit with limits & Disclose and declare & Multiple & Silent & Y & Supports \\
University of Sydney & Australia & Research-intensive & Permit with limits & Multiple & Multiple & Avoid/\allowbreak{}ban & Y & Supports \\
KU Leuven & Belgium & Research-intensive & Integrate/\allowbreak{}AI-expected & Integrate & Open & Silent & N & NA \\
University of British Columbia & Canada & Research-intensive & Devolved to staff & Integrate & Multiple & Silent & N & Mixed \\
University of Toronto (Mississauga) & Canada & Research-intensive & Prohibit & Detect and punish & Secure & Silent & N & Mixed \\
Elsevier & Intl/\allowbreak{}Global & NA (publisher) & Permit with limits & Disclose and declare & Unclear & Silent & Y & Mixed \\
UNESCO & Intl/\allowbreak{}Global & NA (sector body) & Permit with limits & Multiple & Multiple & Indicator only & Y & Supports \\
Hong Kong Metropolitan University & Hong Kong & Teaching-intensive & Disclosure-based & Disclose and declare & Multiple & Silent & Y & Challenges \\
UOW College Hong Kong & Hong Kong & Teaching-intensive & Devolved to staff & Multiple & Multiple & Silent & Y & Supports \\
International Committee of Medical Journal Editors (ICMJE) & Intl/\allowbreak{}Global & NA (professional body) & Permit with limits & Disclose and declare & Unclear & Silent & N & Supports \\
Journal of University Teaching \& Learning Practice (authors: Perkins, Furze, Roe, MacVaugh - British University Vietnam, Deakin University, James Cook University Singapore) & Intl/\allowbreak{}Global & NA (research paper) & Permit with limits & Integrate & Multiple & Avoid/\allowbreak{}ban & Y & Supports \\
Nature Portfolio /\allowbreak{} Springer Nature & Intl/\allowbreak{}Global & NA (publisher) & Permit with limits & Disclose and declare & Hybrid & Silent & Y & Supports \\
Springer Nature & Intl/\allowbreak{}Global & NA (publisher) & Permit with limits & Disclose and declare & Hybrid & Silent & Y & Supports \\
Springer Nature & Intl/\allowbreak{}Global & NA (professional body) & Permit with limits & Disclose and declare & Multiple & Silent & Y & Supports \\
Springer Nature & Intl/\allowbreak{}Global & NA (publisher) & Permit with limits & Disclose and declare & Multiple & Avoid/\allowbreak{}ban & Y & Supports \\
Springer Nature & Intl/\allowbreak{}Global & NA (publisher) & Permit with limits & Disclose and declare & Unclear & Avoid/\allowbreak{}ban & Y & Supports \\
IADT Dun Laoghaire & Ireland & Teaching-intensive & Devolved to staff & Integrate & Hybrid & Silent & N & Supports \\
Nanyang Technological University (Library summary) & Singapore & Research-intensive & Disclosure-based & Disclose and declare & Controlled & Silent & N & Challenges \\
Stellenbosch University (Centre for Teaching and Learning) & South Africa & Research-intensive & Permit with limits & Redesign & Hybrid & Avoid/\allowbreak{}ban & N & Supports \\
University of the Witwatersrand (CLM Faculty) & South Africa & Research-intensive & Permit with limits & Redesign & Hybrid & Avoid/\allowbreak{}ban & Y & Supports \\
Uppsala University & Sweden & Research-intensive & Permit with limits & Integrate & Multiple & Silent & N & Supports \\
ZHAW Zurich University of Applied Sciences & Switzerland & Teaching-intensive & Permit with limits & Multiple & Multiple & Avoid/\allowbreak{}ban & Y & Supports \\
United Arab Emirates University & UAE & Research-intensive & Permit with limits & Multiple & Multiple & Require/\allowbreak{}use & Y & Mixed \\
Arts University Bournemouth (AUB) & UK & Teaching-intensive & Permit with limits & Redesign & Hybrid & Avoid/\allowbreak{}ban & N & Supports \\
Durham University Business School (authors); analyses 24 Russell Group universities & UK & NA (research paper) & Unclear & Multiple & Multiple & Indicator only & N & Mixed \\
Imperial College London & UK & Research-intensive & Devolved to staff & Disclose and declare & Multiple & Silent & N & Supports \\
Imperial College London & UK & Research-intensive & Devolved to staff & Redesign & Unclear & Silent & N & Supports \\
Imperial College London & UK & Research-intensive & Devolved to staff & Multiple & Multiple & Silent & Y & Supports \\
Imperial College London (Centre for Languages, Culture and Communication, CLCC) & UK & Research-intensive & Prohibit & Disclose and declare & Controlled & Silent & Y & Mixed \\
Imperial College London (Educational Development Unit) & UK & Research-intensive & Devolved to staff & Redesign & Multiple & Silent & N & Supports \\
Jisc & UK & NA (grey literature) & Integrate/\allowbreak{}AI-expected & Multiple & Multiple & Silent & Y & Supports \\
King's College London & UK & Research-intensive & Unclear & Unclear & Unclear & Silent & N & NA \\
King's College London & UK & Research-intensive & Permit with limits & Multiple & Multiple & Avoid/\allowbreak{}ban & Y & Mixed \\
King's College London & UK & Research-intensive & Devolved to staff & Multiple & Multiple & Indicator only & Y & Supports \\
King's College London & UK & Research-intensive & Devolved to staff & Multiple & Multiple & Silent & Y & Supports \\
King's College London & UK & Research-intensive & Permit with limits & Multiple & Multiple & Silent & Y & Mixed \\
King's College London & UK & Research-intensive & Permit with limits & Disclose and declare & Hybrid & Avoid/\allowbreak{}ban & Y & Supports \\
King’s College London & UK & Research-intensive & Permit with limits & Multiple & Multiple & Avoid/\allowbreak{}ban & N & Supports \\
King’s College London & UK & Research-intensive & Integrate/\allowbreak{}AI-expected & Integrate & AI-integrated & Silent & Y & Supports \\
King’s College London (cross-institutional collaboration with LSE and University of Southampton) & UK & Research-intensive & Unclear & Integrate & Unclear & Avoid/\allowbreak{}ban & N & Supports \\
QAA (Quality Assurance Agency for Higher Education) & UK & NA (sector body) & Unclear & Multiple & Multiple & Avoid/\allowbreak{}ban & N & Supports \\
Quality Assurance Agency for Higher Education (QAA) & UK & NA (sector body) & Unclear & Unclear & Unclear & Silent & N & Neutral \\
Tao, Ryan, Bialy \& Zhang (Lancaster University, University of Manchester, University of Liverpool) — Research Square preprint & UK & NA (research paper) & Unclear & Multiple & Multiple & Silent & Y & Mixed \\
University College London (UCL) & UK & Research-intensive & Devolved to staff & Multiple & Multiple & Avoid/\allowbreak{}ban & Y & Supports \\
University of Oxford & UK & Research-intensive & Devolved to staff & Multiple & Multiple & Indicator only & Y & Mixed \\
University of Oxford & UK & Research-intensive & Prohibit & Multiple & Multiple & Silent & N & Mixed \\
University of Oxford & UK & Research-intensive & Permit with limits & Disclose and declare & Unclear & Silent & Y & Supports \\
Taylor \& Francis & UK & NA (publisher) & Permit with limits & Disclose and declare & Unclear & Silent & N & Supports \\
Committee on Publication Ethics (COPE) & UK & NA (professional body) & Disclosure-based & Disclose and declare & Unclear & Silent & N & Supports \\
Science Journals /\allowbreak{} American Association for the Advancement of Science (AAAS) & Intl/\allowbreak{}Global & NA (publisher) & Permit with limits & Disclose and declare & Unclear & Silent & N & Supports \\
Carnegie Mellon University (Eberly Center) & USA & Research-intensive & Devolved to staff & Multiple & Multiple & Silent & Y & Mixed \\
Harvard University (FAS, Office of Undergraduate Education) & USA & Research-intensive & Devolved to staff & Multiple & Multiple & Lockdown/\allowbreak{}prevent & Y & Mixed \\
Massachusetts Institute of Technology (Ad Hoc Committee on AI Use in Teaching, Learning, and Research Training) & USA & Research-intensive & Devolved to staff & Redesign & Multiple & Avoid/\allowbreak{}ban & Y & Supports \\
Stanford University & USA & Research-intensive & Prohibit & Detect and punish & Secure & Indicator only & N & Mixed \\
\end{longtable}
\end{landscape}

\section{Literature corpus coding}
\begin{landscape}
\scriptsize
\setlength{\tabcolsep}{4pt}
\renewcommand{\arraystretch}{1.1}
\begin{longtable}{@{}>{\raggedright\arraybackslash}p{5.5cm} >{\raggedright\arraybackslash}p{2.1cm} >{\raggedright\arraybackslash}p{2.9cm} >{\raggedright\arraybackslash}p{1.5cm} >{\raggedright\arraybackslash}p{1.7cm} >{\raggedright\arraybackslash}p{1.7cm}@{}}
\caption{Literature corpus: all 41 sources behind the codings summarised in Table~\ref{tab:litcounts}, by source type, research-question relevance, evidence strength, proposition direction and AI capability zone. Source types total 20 conceptual, 11 empirical, 5 reviews, 3 commentaries and 2 reports; proposition directions total 22 supports, 8 mixed, 2 neutral and 9 not applicable, the last including the six methodology references (Methods). Capability zones total 5 AI-strong, 1 AI-weaker, 17 mixed and 18 not applicable. The two EDUCAUSE Horizon Reports are typed as recorded in the source coding sheet (2025 as conceptual, 2026 as commentary); both are sector trend reports.}\label{tab:litmatrix}\\
\toprule
\textbf{Source} & \textbf{Type} & \textbf{RQ relevance} & \textbf{Strength} & \textbf{Direction} & \textbf{AI zone} \\
\midrule
\endfirsthead
\multicolumn{6}{@{}l}{\footnotesize\itshape Table \ref{tab:litmatrix} continued}\\
\toprule
\textbf{Source} & \textbf{Type} & \textbf{RQ relevance} & \textbf{Strength} & \textbf{Direction} & \textbf{AI zone} \\
\midrule
\endhead
\midrule \multicolumn{6}{r@{}}{\footnotesize continued on next page}\\
\endfoot
\bottomrule
\endlastfoot
Arksey and O'Malley (2005) & Conceptual & Methods & Low & NA & NA \\
Bearman et al. (2016) & Empirical & RQ3 & Medium & NA & NA \\
Biggs (1996) & Conceptual & RQ3, RQ5 & Medium & Supports & AI-weaker \\
Boud and Falchikov (2006) & Conceptual & RQ3, RQ5 & Medium & Neutral & NA \\
Bowen (2009) & Conceptual & Methods & Low & NA & NA \\
Braun and Clarke (2006) & Conceptual & Methods & Low & NA & NA \\
Bretag (Ed.) (2016) & Conceptual & RQ1, RQ2, RQ3, RQ4, RQ5 & Medium & Supports & NA \\
Bridgeman and Liu (2025) & Commentary & RQ2, RQ3, RQ4 & Medium & Supports & Mixed \\
Chan (2023) & Empirical & RQ1, RQ2, RQ3, RQ4 & Medium & Supports & Mixed \\
Chan and Hu (2023) & Empirical & RQ1, RQ2, RQ5 & Medium & Supports & Mixed \\
Corbin et al. (2025) & Conceptual & RQ2, RQ3, RQ4 & Medium & Supports & AI-strong \\
Cotton et al. (2024) & Conceptual & RQ1, RQ2, RQ3 & Medium & Mixed & AI-strong \\
Dawson (2021) & Conceptual & RQ1, RQ2, RQ3, RQ4, RQ5 & Medium & Mixed & Mixed \\
Digital Education Council (2026) & Empirical & RQ1, RQ2, RQ3, RQ4, RQ5 & Medium & Mixed & Mixed \\
Digital Education Council and Pearson (2025) & Review & RQ3, RQ5, RQ2 & Medium & Supports & Mixed \\
European Union (AI Act) (2024) & Conceptual & RQ1, RQ2, RQ4 & High & Neutral & NA \\
Freeman (HEPI) (2025) & Empirical & RQ1, RQ2, RQ4 & Medium & Supports & Mixed \\
Ivković et al. (2025) & Conceptual & RQ1, RQ2, RQ3, RQ5 & Medium & Mixed & AI-strong \\
Jiao et al. (2024) & Review & RQ2, RQ3, RQ4 & High & Mixed & NA \\
Jiao et al. (2025) & Review & RQ2 & Medium & NA & NA \\
Kasneci et al. (2023) & Conceptual & RQ1, RQ2, RQ3, RQ4, RQ5 & Medium & Supports & Mixed \\
Lancaster and Clarke (2008) & Conceptual & RQ1, RQ2, RQ3 & Medium & Supports & NA \\
Lancaster and Clarke (2017) & Empirical & RQ1, RQ3, RQ4 & Medium & Mixed & NA \\
Levac et al. (2010) & Conceptual & Methods & Low & NA & NA \\
Li et al. (2025) & Review & RQ1, RQ4 & Low & NA & NA \\
Liang et al. (2023) & Empirical & RQ1, RQ4 & High & Supports & Mixed \\
Maranhão et al. (FDUSP) (2024) & Report & RQ1, RQ2, RQ4 & Medium & Mixed & NA \\
Mollick (2023) & Commentary & RQ1, RQ3, RQ5 & Low & Supports & AI-strong \\
Moorhouse, Yeo \& Wan (2023) & Review & RQ2, RQ3 & High & Supports & Mixed \\
Nikolić and Basta Nikolić (2026) & Conceptual & RQ1, RQ3, RQ4, RQ5 & Medium & Supports & Mixed \\
OECD (2026) & Report & RQ1, RQ2, RQ3, RQ4, RQ5 & Medium & Supports & Mixed \\
Perkins et al. (2024) & Conceptual & RQ2, RQ3, RQ4, RQ5 & Medium & Supports & Mixed \\
Robert et al. (2025) & Conceptual & RQ2, RQ3, RQ4, RQ5 & Medium & Supports & Mixed \\
Robert et al. (2026) & Commentary & RQ1, RQ2, RQ3, RQ5 & Medium & Supports & Mixed \\
Schuwirth and van der Vleuten (2011) & Conceptual & RQ3, RQ5 & Medium & Supports & NA \\
Stephenson and Armstrong (2026) & Empirical & RQ1, RQ2, RQ3 & Medium & Mixed & Mixed \\
Thomas and Harden (2008) & Empirical & Methods & Low & NA & NA \\
Tricco et al. (2018) & Empirical & Methods & Low & NA & NA \\
Univ. of Wisconsin (PROTECT) (2026) & Conceptual & RQ1, RQ3, RQ4 & Low & Supports & Mixed \\
Weber-Wulff et al. (2023) & Empirical & RQ1, RQ2, RQ4, RQ5 & High & Supports & AI-strong \\
Wiggins (1990) & Conceptual & RQ3, RQ4, RQ5 & Medium & Supports & NA \\
\end{longtable}
\end{landscape}

\bibliography{references}

\end{document}